\documentstyle[emulateapj,onecolfloat]{article}

\begin{document}
\twocolumn[
\submitted{AJ, accepted}
\title{Colors, magnitudes and velocity dispersions in early-type galaxies: 
       Implications for galaxy ages and metallicities}
\author{Mariangela Bernardi$^{1,2}$, Ravi K. Sheth$^{1,2}$, 
        Robert C. Nichol$^{3}$, D. P. Schneider$^4$, and J. Brinkmann$^5$}

\begin{abstract}
We present an analysis of the color-magnitude-velocity dispersion relation 
for a sample of 39320 early-type galaxies within the Sloan Digital Sky Survey. 
We demonstrate that the color-magnitude relation is entirely a consequence of 
the fact that both the luminosities and colors of these galaxies are correlated 
with stellar velocity dispersions.  Previous studies of the color-magnitude 
relation over a range of redshifts suggest that the luminosity of an 
early-type galaxy is an indicator of its metallicity, whereas residuals 
in color from the relation are indicators of the luminosity-weighted age of 
its stars.  
We show that this, when combined with our finding that velocity 
dispersion plays a crucial role, has a number of interesting implications.  
First, galaxies with large velocity dispersions tend to be older 
(i.e., they scatter redward of the color-magnitude relation).  
Similarly, galaxies with large dynamical mass estimates also tend to 
be older.  In addition, at fixed luminosity, galaxies which are smaller, 
or have larger velocity dispersions, or are more massive, tend to be 
older.  
Second, models in which galaxies with the largest velocity dispersions 
are also the most metal poor are difficult to reconcile with our data.  
However, at fixed velocity dispersion, galaxies have a range of ages 
and metallicities:  the older galaxies have smaller metallicities, and 
vice-versa.  
Finally, a plot of velocity dispersion versus luminosity can be used 
as an age indicator:  lines of constant age run parallel to the 
correlation between velocity dispersion and luminosity.  
\end{abstract}
]

\footnotetext[1] {Department of Physics and Astronomy, 
                 University of Pittsburgh, Pittsburgh, PA 15620}
\footnotetext[2] {Department of Physics and Astronomy, 
                 University of Pennsylvania, Philadelphia, PA 15104}
\footnotetext[3] {Institute of Cosmology and Gravitation (ICG), Mercantile House, Hampshire Terrace, University of Portsmouth, Portsmouth, PO1 2EG, UK}
\footnotetext[4] {Department of Astronomy and Astrophysics, The Pennsylvania State University, University Park, PA 16802}
\footnotetext[5] {Apache Point Observatory, 2001 Apache Point Road, P.O. Box 59, Sunspot, NM 88349-0059}

\section{Introduction}
It has long been known that early-type galaxies tend to be the reddest 
galaxies, and that the more luminous the galaxy, the redder its color 
(e.g. Sandage \& Viswanathan 1978a,b; Bower, Lucey \& Ellis 1992a,b).  
The tightness of the correlation between color and magnitude has 
been used to constrain models of how early-type galaxies formed.  
However, a luminous galaxy may appear red either because its stars are 
older, or because, although its stars are younger, they are more metal rich.  
This has complicated the constraints one can place on galaxy formation 
models:  should the models produce luminous metal-rich 
galaxies, or luminous old galaxies?  
To illustrate, Appendix~\ref{bc2003} shows the color-magnitude relation 
associated with the recent stellar population synthesis models of 
Bruzual \& Charlot (2003).  

The colors and luminosities of older galaxies are expected to evolve 
more slowly than those of younger ones.  Therefore, if age is changing 
along the color-magnitude sequence (e.g. if the more luminous galaxies 
are older), one would expect the differential evolution of the older 
and younger populations along the sequence to manifest as a change in 
the slope of the color-magnitude relation with redshift.  
Recent measurements have shown that the relation was already in place 
at redshifts of order unity, and that its slope appears to be unchanged 
from its value locally (e.g. Kodama et al. 1998; Blakeslee et al. 2003).  
If the high redshift population does indeed represent the local low 
redshift population in its youth, then the fact that the slope has 
evolved little argues against a large age-spread along the relation.  
In this case, the color-magnitude relation is caused primarily by 
a correlation between metallicity and luminosity (Kodama et al. 1998).  
(This argument becomes weaker if the redshift at which the population 
formed the bulk of its stars is large.)  

In this framework, the scatter around the color-magnitude relation is 
usually attributed to the effects of age.  It is not entirely obvious, 
however, that this can be correct.  This is because differential 
evolution must make the scatter around the mean relation larger at 
higher lookback times.  Therefore, the fact that the color-magnitude 
relation is well defined at redshifts of order unity can be translated 
into a constraint on the mix of ages present at redshift zero.  For this 
reason, it would be interesting to quantify how the scatter around the 
mean relation evolves.  (Blakeslee et al. 2003 find little evolution 
in the slope and scatter of the relation, using cluster early-types out 
to redshifts of order unity.)  

It is well known that luminosity also correlates with velocity 
dispersion (Poveda 1961; Faber \& Jackson 1976), so it is natural to 
ask if velocity dispersion is also tightly coupled to metallicity.  
The tightness of the correlation between Mg$_2$ and $\sigma$ 
(e.g. Bernardi et al. 1998; Colless et al. 1999), is thought to 
be a consequence of variations in age and metallicity which conspire 
to keep the observed correlation tight (e.g. Kuntschner et al. 2001).  
In this view, old galaxies with a given $\sigma$ tend to be metal 
poor, whereas younger galaxies with the same $\sigma$ are metal rich 
(e.g. Worthey 1994; Trager et al. 2000b).  

Thus, recent work suggests that residuals from the color-magnitude 
relation are age indicators, whereas studies of chemical abundances 
suggest that velocity dispersion is a combination of both age and 
metallicity.  One of the goals of the present work is to see if these 
conclusions are consistent with one-another.  
We do this by studying the joint distribution of color, luminosity 
and velocity dispersion in a sample of 39320 early-type galaxies 
drawn from the Sloan Digital Sky Survey (York et al. 2000; 
Stoughton et al. 2002; Abazajian et al. 2003) database (hereafter SDSS).  

Section~\ref{sample} describes our sample.  Section~\ref{cms} 
presents the color-magnitude and color-$\sigma$ relations, and 
demonstrates that color-$\sigma$ and magnitude-$\sigma$ are the primary 
correlations.  A simple model is introduced which illustrates clearly 
what our measurements imply for the relations between age, metallicity 
and velocity dispersion.  The mathematics associated with this model is 
in Appendix~\ref{algebra}.  
Section~\ref{discuss} uses the model to derive a number of consequences 
of the observed color-magnitude-$\sigma$ relations, and summarizes our 
findings.  

Throughout, we assume that $H_0=70$~km~s$^{-1}$~Mpc$^{-1}$ in a 
universe with $\Omega_0=0.3$ which is spatially flat.  

\begin{figure}[t]
 \centering
 \epsfxsize=\hsize\epsffile{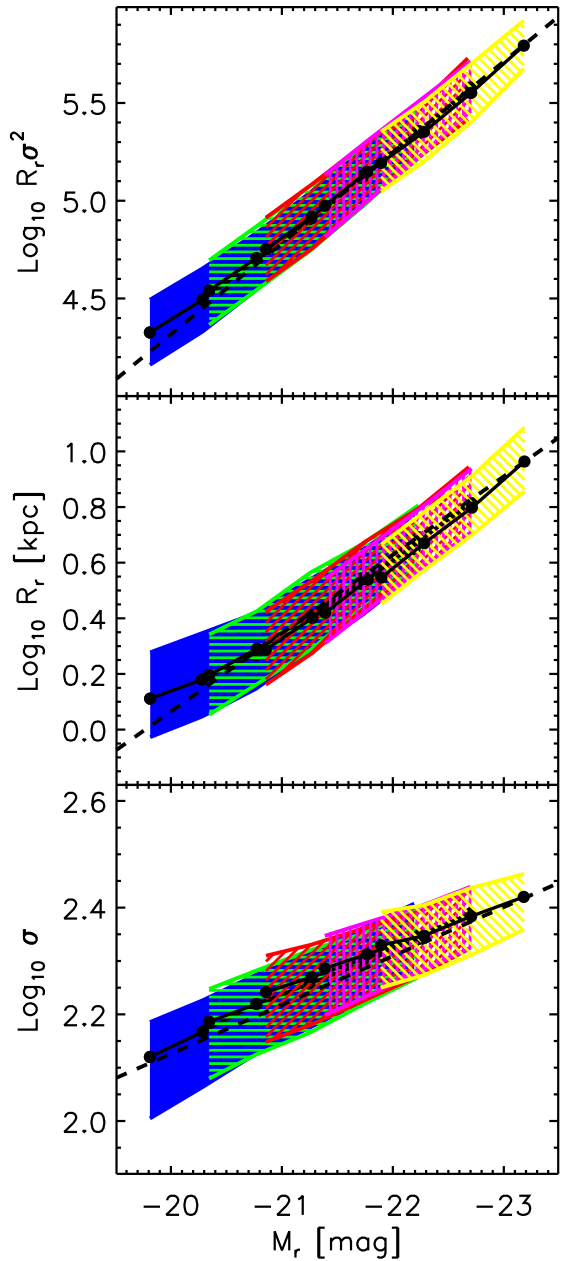}
 \caption{Luminosity correlates with both velocity dispersion (bottom) 
          and size (middle), but the tightest correlation is with 
          $R_r\sigma^2 \propto$~mass (top).  Dashed lines show 
          the correlations summarized in Table~\ref{MLcov}.  
          Luminosities have been corrected for evolution by 0.85$z$ 
          (see text for details).}
 \label{lvrm}
\end{figure}

\section{The SDSS early-type galaxy sample}\label{sample}
For our analysis we used galaxies selected from the Sloan Digital Sky 
Survey (SDSS) database.
See York et al. (2000) for a technical summary of the SDSS project; 
Stoughton et al. (2002) for a description of the Early Data Release; 
Abazajian et al. (2003) et al. for a description of DR1, the First Data 
Release; Gunn et al. (1998) for details about the camera; 
Fukugita et al. (1996), Hogg et al. (2001) and Smith et al. (2002) 
for details of the photometric system and calibration; 
Lupton et al. (2001) for a discussion of the photometric data reduction 
pipeline; Pier et al. (2003) for the astrometric calibrations; 
Blanton et al. (2003) for details of the 
tiling algorithm; Strauss et al. (2002) and Eisenstein et al. (2001)
for details of the target selection. 

\begin{table*}
\centering
\caption[]{Maximum-likelihood estimates, in the SDSS $r$ band, of the 
joint distribution of luminosities, sizes and velocity dispersions.  
The mean values of the variables at redshift $z$, $M_*-Qz$, $R_*$, $V_*$, 
and the coefficients of the various pairwise correlations between the 
variables are shown.  The coefficient which differs most from Table~1 
of Bernardi et al. (2003b) is $R_*$.\\}
\begin{tabular}{ccccccccccc}
\tableline 
 Band & $M_*$ & $\sigma_{MM}$ & $R_*$ & $\sigma_{RR}$ &
 $V_*$ & $\sigma_{VV}$ & $\xi_{RM}$ & $\xi_{VM}$ & $\xi_{RV}$ & Q\\
 \hline\\
 $r$  & $-21.025$ & 0.841 & 0.354 & 0.263 & 2.220 & 0.111 & $-0.900$ &
 $-0.774$ & 0.550 & 0.85 \\
\tableline
\end{tabular}
 \label{MLcov}
\end{table*}

Our early-type galaxy sample is selected similarly to that of 
Bernardi et al. (2003a), with some minor changes:  
We selected all objects targeted as galaxies and having 
de-reddened Petrosian apparent magnitude $14.5 \le r_{\rm Pet}\le 17.75$.  
To extract a sample of early-type galaxies we then chose the subset 
with the spectroscopic parameter {\tt eclass < 0} ({\tt eclass} classifies 
the spectral type based on a Principal Component Analysis), and the 
photometric parameter {\tt fracDev$_r$ > 0.8}. (The parameter {\tt fracDev} is 
a seeing-corrected indicator of morphology.  It is obtained by taking 
the best fit exponential and de Vaucouleurs fits to the surface 
brightness profile, finding the linear combination of the two that 
best-fits the image, and storing the fraction contributed by the 
de Vaucouleurs fit.) We removed galaxies with problems in the 
spectra (using the {\tt zStatus} and {\tt zWarning} flags).
From this subsample, we finally chose those objects for which the 
spectroscopic pipeline had measured velocity dispersions (meaning that 
the signal-to-noise ratio in pixels between the restframe wavelengths 
4200\AA\ and 5800\AA\ is S/N $>10$).  These selection criteria produced 
a sample of 39320 objects with photometric parameters output by version 
${\tt V5.4}$ of the SDSS photometric pipeline and ${\tt V.23}$ 
reductions of the spectroscopic pipeline (i.e. ${\tt v4\_10\_2}$ of 
${\tt idlspec2d}$ and ${\tt v5\_9\_4}$ of ${\tt spectro1d}$). 

The main quantities we use in this paper are magnitudes, sizes, colors, 
redshifts and velocity dispersions; 
the first three are output from the SDSS photometric pipeline, and 
the final two from the SDSS spectroscopic pipeline.  
The photometry is accurate to 0.02~mags, and the velocity dispersions 
to about $8$\% (see Bernardi et al. 2003a).  
We work with de Vaucouleurs (1948) magnitudes and sizes, and SDSS 
{\tt model} colors throughout:  these are colors measured within an 
aperture which scales with the de Vaucouleurs half-light radius in the 
$r$-band.  

\begin{figure*}[t]
 \centering
 \epsfxsize=\hsize\epsffile{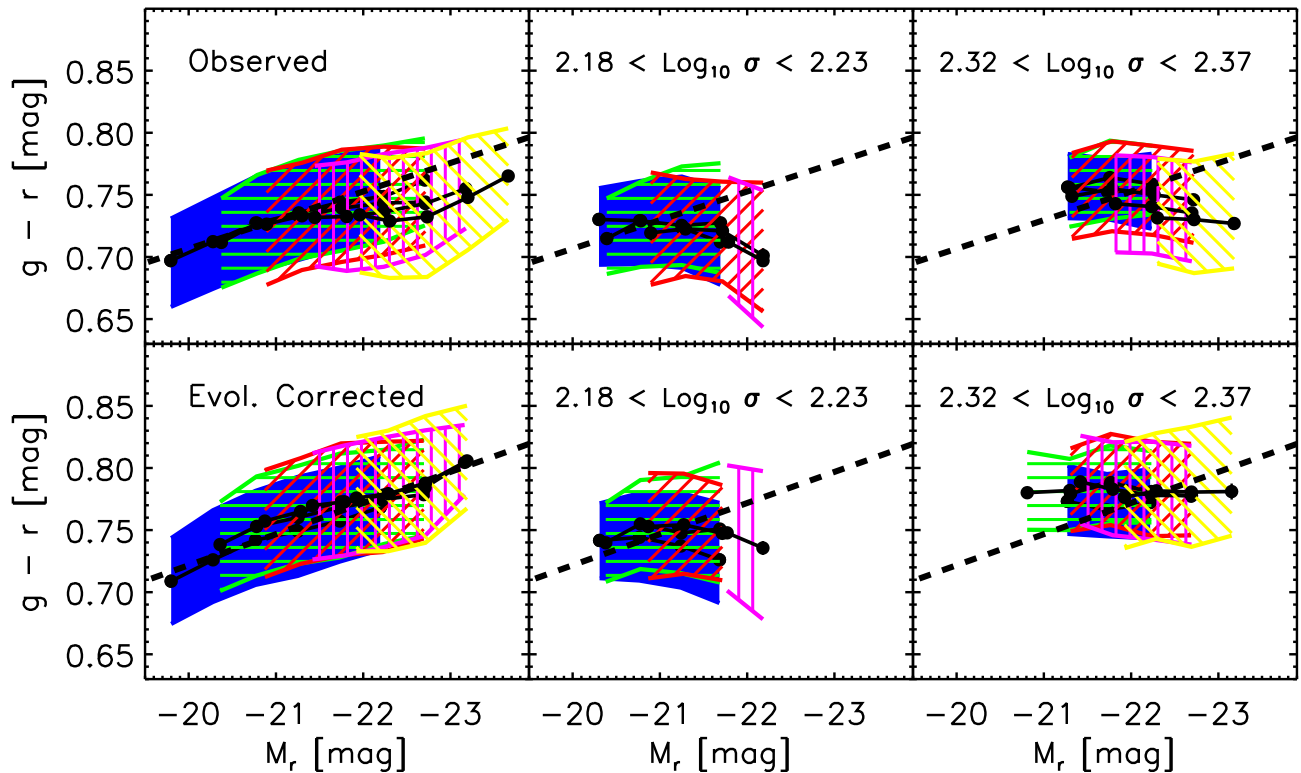}
 \caption{Color-magnitude relation in the sample.  Top series of panels 
          show the raw measurement, and bottom panels show the result 
          of making magnitudes fainter by $0.85z$~mag and colors redder 
          by $0.3z$~mag to account for evolution.  Both on the top and 
          bottom, the left-most panel shows the color-magnitude relation 
          in the full sample, and the other two panels show the relation 
          in small bins in velocity dispersion.  
          Clearly, at fixed velocity dispersion, there is no 
          relation between color and magnitude.  }
 \label{cmag}
\end{figure*}

The most important reason for analyzing a new sample (other than size) 
is that the old photometric reductions output by the SDSS pipeline 
{\tt photo} were incorrect (see documentation on DR2, the Second Data 
Release).  
A comparison of the properties of objects for which old and new 
photometric reductions are available shows that the new corrected 
photometry has made most magnitudes slightly fainter ($\sim 0.13$~mags), 
and most half-light radii smaller ($\sim$10\%).  
It is customary to report velocity dispersions at some fraction 
(we use 1/8) of the half-light radius.  The half-light radii of the 
galaxies which enter our sample are typically about 2~arcsecs, 
approximately independent of redshift, whereas the SDSS fiber used to 
measure the spectrum from which the velocity dispersion is estimated 
has a diameter of 3~arcsec.  Since the velocity dispersions of early-type 
galaxies are known to increase towards the center (following J{\o}rgensen 
et al. 1995 we assume the scaling is $\propto (r/r_e)^{0.04}$, 
where $r_e$ is the half-light radius), all measured velocity dispersions 
are `aperture corrected', and it is these which are usually reported 
(the mean correction is 7\%).  
Although the measured velocity dispersions have not changed, 
the new photometry has changed the half-light radius.  
Hence, aperture corrected velocity dispersions differ from those 
associated with the old photometry (i.e., those reported in 
Bernardi et al. 2003a) by less than one percent 
(the new values are larger by a factor of $1.1^{0.04}$).  

\begin{table}
\centering
\caption[]{Maximum-likelihood estimates of the joint distribution of 
color, $r$-band magnitude and velocity dispersion and its evolution.  
At redshift $z$, the mean values are $C_*-Pz$,  and the covariances 
are 
  $\langle(C-C_*)(M-M_*)\rangle = \sigma_{CC}\sigma_{MM}\,\xi_{CM}$, 
and similarly for $\xi_{CV}$ ($M_*$ and $V_*$ are given in 
Table~\ref{MLcov}).\\}
\begin{tabular}{cccccc}
\tableline 
Color & $C_*$ & $\sigma_{CC}$ & $\xi_{CM}$ & $\xi_{CV}$ & $P$ \\
%     & mag & mag & dex & dex & dex & dex & & & \\
\hline\\
$g-r$ & 0.736 & 0.057 & $-0.361$ & 0.516 & 0.30\\
\tableline
\end{tabular}
\label{MLcmag}
\end{table}

Bernardi et al. (2003b) showed that the luminosities in their sample 
($\sim 9000$ galaxies with $z\le 0.3$) evolve:  
$M_*(z) = M_*(0) - 0.85z$, where $M_*(z)$ denotes the mean absolute 
magnitude at redshift $z$.  We find similar evolution in the new sample:  
the main difference is that the new photometric reductions make $M_*$ 
fainter by about 0.125~mags.  
The evolution is consistent with that of a passively aging population.  
Thus, in the analysis which follows, we 
will be careful to separate trends with redshift that are due to 
the magnitude limit of the sample, from trends that are due to evolution.  
Many of the following figures present measurements made in a 
series of narrow redshift bins:  $0.02\le z<0.07$, $0.07\le z<0.09$, 
$0.09\le z < 0.12$, $0.12\le z< 0.15$, and $0.15\le z<0.2$.  
In all of these figures, symbols show the median value of the variable 
on the y-axis as a function of the observable on the x-axis, error bars 
show the rms error of the median, and hashed regions indicate where 
the central 68\% of the sample in each redshift bin lies.  

Figure~\ref{lvrm} displays correlations between luminosity, size 
$R_r$, and velocity dispersion $\sigma$ in this sample.  (The 
half-light radius in kpc, $R_r$, is defined similarly to the quantity 
called $R_o$ by Bernardi et al. 2003a, for the $r$-band.)  The dashed 
lines show fits to these correlations, summarized in Table~\ref{MLcov}.  
In all cases, luminosities have been corrected for evolution by 
0.85$z$.  In the Table, 
$M_*$ and $\sigma_M$ are the mean and rms values of a Gaussian fit 
to the evolution-corrected $r$-band luminosity function, 
$Q$ quantifies the rate of luminosity evolution, and 
$R_*$ and $V_*$ are the values of $\log_{10}$(size/$h^{-1}$kpc) and 
$\log_{10}$(velocity dispersion/km~s$^{-1}$) for an $L_*$ galaxy.  
In general, these values are all qualitatively similar to those 
reported in Bernardi et al. (2003b), but there are quantitative 
differences, most noticably in the mean size $R_*$, which is smaller 
by about ten percent.  
Table~\ref{MLcov} also reports values of a number of pairwise 
correlations in the following format.  
Given a pair of observables $X$ and $Y$, 
$\langle Y-Y_*|X\rangle/\sigma_{YY} = \xi_{XY}(X-X_*)/\sigma_{XX}$.  
The results of a similar analysis of the correlations between color, 
magnitude and velocity dispersion is given in Table~\ref{MLcmag}.

\section{Color-magnitude and color-$\sigma$ relations}\label{cms}

Figure~\ref{cmag} shows the color-magnitude relation in our sample.  
The top series of panels shows the raw measurement, and bottom panels 
show the result of accounting for evolution by adding $0.85z$~mags 
to the magnitudes, and adding $0.3z$~mags to the colors (see 
Table~\ref{MLcmag}).  On both tiers, the left-most panel shows the 
color-magnitude relation in the full sample, and the other two panels 
show the relation in small bins in velocity dispersion.  
The figure makes two important points:  
(i) at fixed velocity dispersion, there is no relation between 
    color and magnitude, and 
(ii) the slope of the relation in the leftmost panel is approximately 
     the same in all redshift bins.  

We also find no evidence that the scatter around the mean relations 
changes with redshift, but placing a precise limit of the evolution is 
difficult, because the expected signal is small, so the answer is 
sensitive to uncertainties in the SDSS photometry (0.02~mags) and the 
$k$-correction.  
Specifically, because we $k$-correct following the procedure described in 
Bernardi et al. (2003a), all galaxies at the same redshift are assigned 
the same $k$-correction.  This leads to a small amount of scatter 
(less than 0.02~mags) in our absolute magnitudes and colors which may 
well depend on redshift.

\begin{figure*}[t]
 \centering
 \epsfxsize=\hsize\epsffile{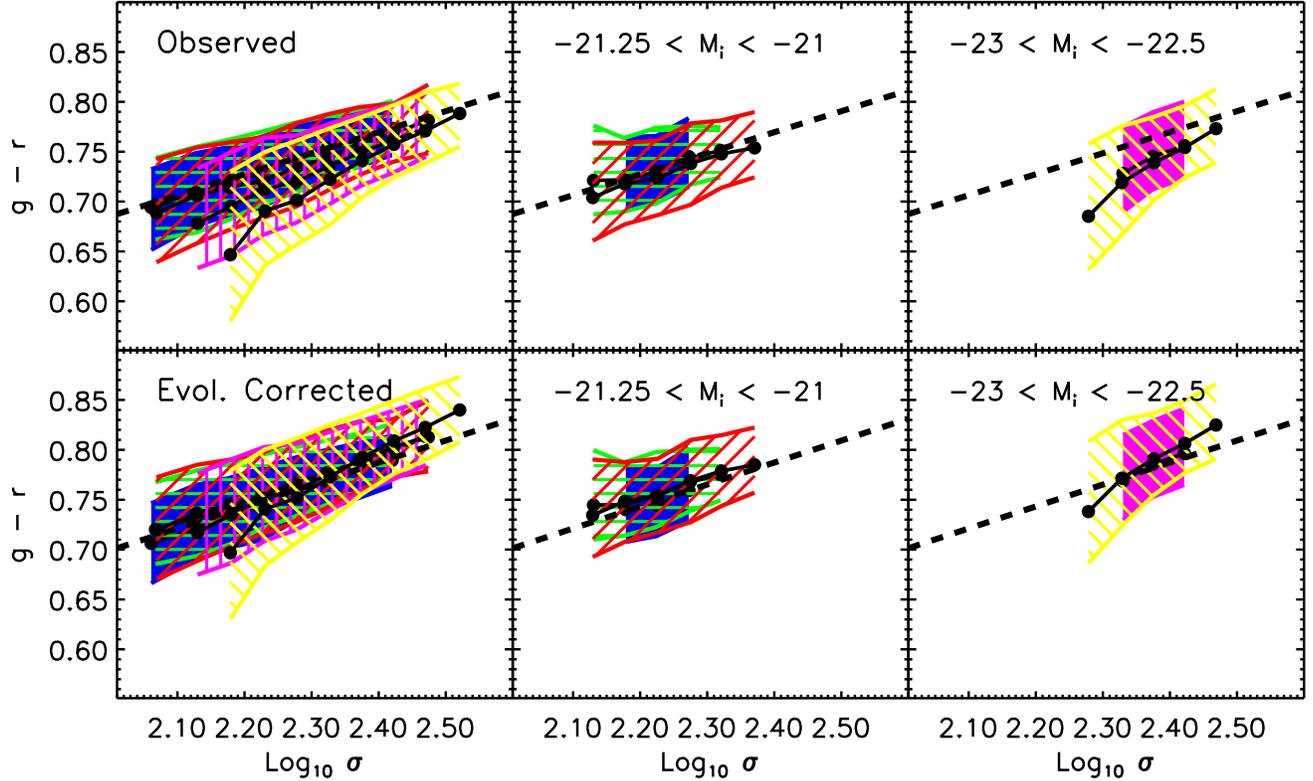}
 \caption{Color-$\sigma$ relation in the sample.  Top series of panels 
          show the raw measurement, and bottom panels show the result 
          of accounting for evolution by adding $0.3z$~mags to the 
          colors, as was required to model the evolution of the color 
          magnitude relation.  Both on the top and bottom, the 
          left-most panel shows the color-$\sigma$ relation 
          in the full sample, and the other two panels show the relation 
          in small bins in magnitude.  
          Clearly, at fixed redshift, the correlation between color 
          and $\sigma$ is the same, whatever the magnitude.  }
 \label{csig}
\end{figure*}

Figure~\ref{csig} shows the color-$\sigma$ relation in the sample.  
The format is the same as for the previous figure:  
top panels show the raw measurement, and bottom panels show the result 
of accounting for evolution by adding $0.3z$~mags to the colors, as was 
required to model the evolution of the color magnitude relation.  
The left-most panels show the color-$\sigma$ relation in the full 
sample, and the other two panels show the relation for galaxies which 
are restricted to a narrow range in magnitude.  
Clearly, at fixed redshift, the correlation between color and $\sigma$ 
is the same for all magnitude bins.

Comparison of the color-$\sigma$ relation in the different redshift 
bins suggests that the relation is steepening slightly with redshift:  
galaxies with small velocity dispersions are evolving slightly more 
rapidly.  Because we are no longer making measurements at fixed 
magnitude, one may wonder if the slope of the color-$\sigma$ relation 
is affected by the magnitude-limited selection.  
Appendix~\ref{algebra} develops a simple model which demonstrates 
that this is not the case.  
We have also checked that the slope change does not depend 
strongly on our choice of $k$-correction.  With the current sample 
size, this steeping is barely significant: it will be interesting to 
see if this steepening persists when the sample is larger.

\begin{figure}[b]
 \centering
 \epsfxsize=\hsize\epsffile{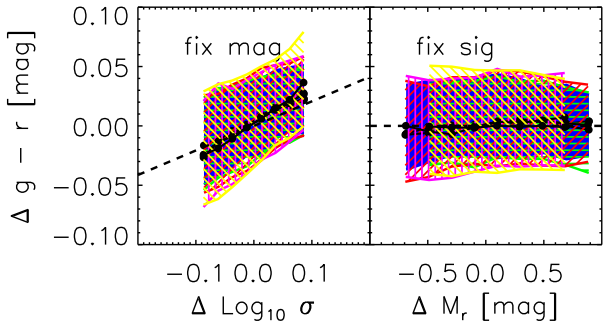}
 \caption{Residuals from the color-magnitude relation correlate with 
          residuals from the $\sigma$-magnitude relation (left) 
          whereas residuals from the color-$\sigma$ relations do not 
          correlate with residuals from the magnitude-$\sigma$ relation 
          (right).  Dashed lines show the relations from our model,  
          equation~(\ref{basic}), obtained by inserting the values 
          from Tables~\ref{MLcov} and~\ref{MLcmag} in 
          equations~(\ref{RcmRvm}) and~(\ref{RmvRcv}), respectively.}
 \label{cvmresids}
\end{figure}

Together, Figures~\ref{cmag} and~\ref{csig} indicate that the 
color-magnitude relation is entirely a consequence of the 
color-$\sigma$ and magnitude-$\sigma$ relations.  In particular, the 
joint distribution of color, magnitude and velocity dispersion is:
\begin{equation}
 p({\rm color},{\rm mag},\sigma) = p(\sigma)\,p({\rm mag}|\sigma)\,
                                   p({\rm color}|\sigma).
 \label{basic}
\end{equation}
Appendix~\ref{algebra} demonstrates that if this expression is correct, 
then residuals from the magnitude-$\sigma$ relation should not correlate 
with residuals from the color-$\sigma$ relation.  This is shown in 
Figure~\ref{cvmresids}:  
$M - \langle M|\sigma\rangle$ does not correlate with 
${\rm color} - \langle {\rm color}|\sigma\rangle$, but 
$\sigma - \langle\sigma|M\rangle$ does correlate with 
${\rm color} - \langle {\rm color}|M\rangle$.  
Moreover, residuals from the color-$\sigma$ relation should not 
correlate with magnitude, whereas residuals from the color-magnitude 
relation should correlate with velocity dispersion.  
The two panels in Figure~\ref{agesigma} show that this is indeed the 
case.  Note that the slope of the correlation between 
${\rm color} - \langle {\rm color}|M\rangle$ and $\sigma$ is not as 
steep as for the color-$\sigma$ relation itself.  In effect, this 
is because $\langle {\rm color}|M\rangle$ correlates with $\sigma$ 
(because $M$ itself correlates with $\sigma$).  

The bottom panel of Figure~\ref{agesigma} shows that the correlation 
between color-magnitude residuals (i.e., age) and velocity dispersion 
steepens systematically with redshift.  Although it is tempting to 
conclude that this is evidence of differential evolution, the analysis 
in Appendix~\ref{algebra} shows that it is, in fact, a selection effect 
(see text between equations~\ref{RcmV} and~\ref{RcmRcv}). 

\section{Interpretation and discussion}\label{discuss}
\begin{figure}[t]
 \centering
 \epsfxsize=0.8\hsize\epsffile{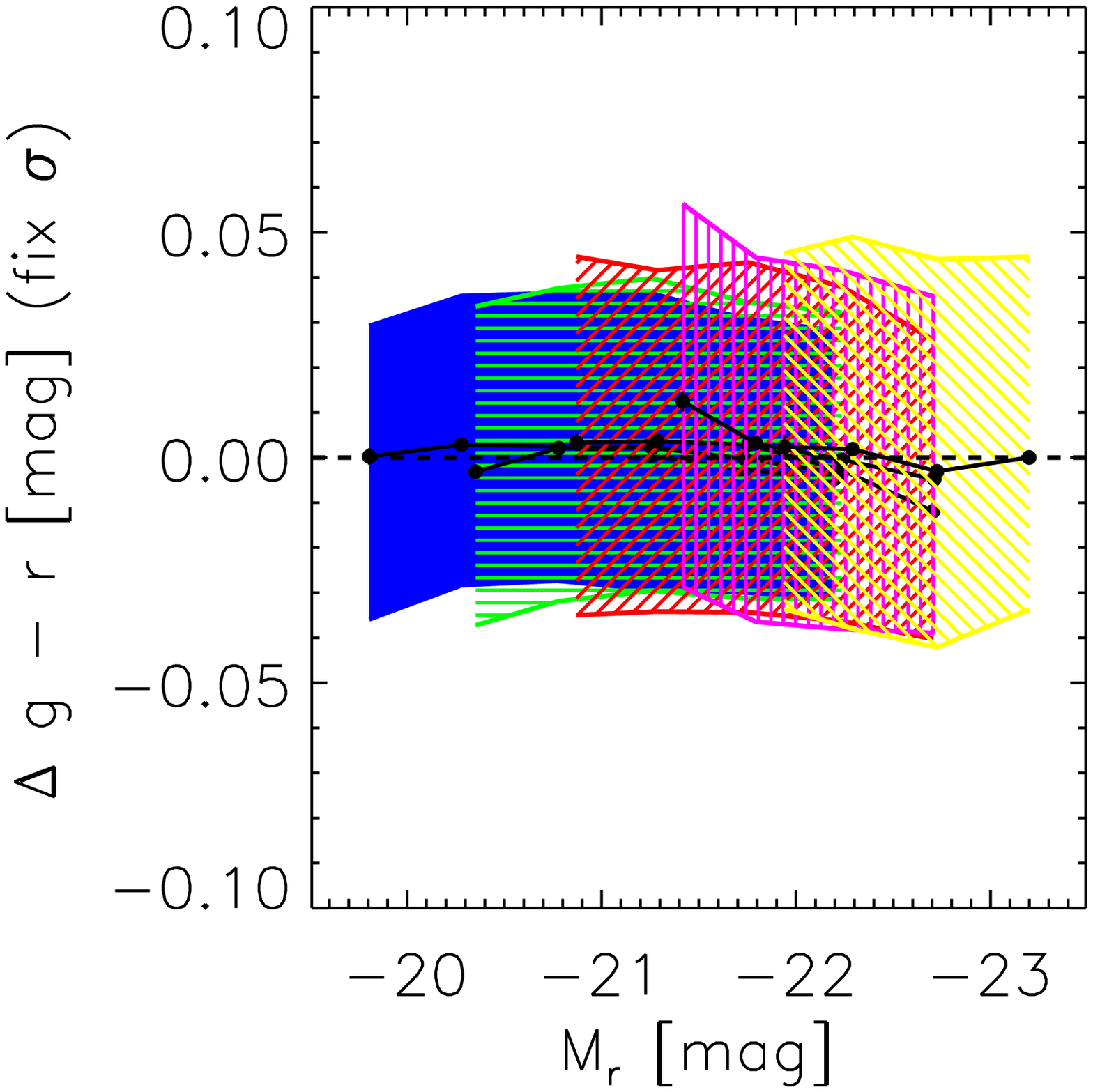}
 \epsfxsize=0.8\hsize\epsffile{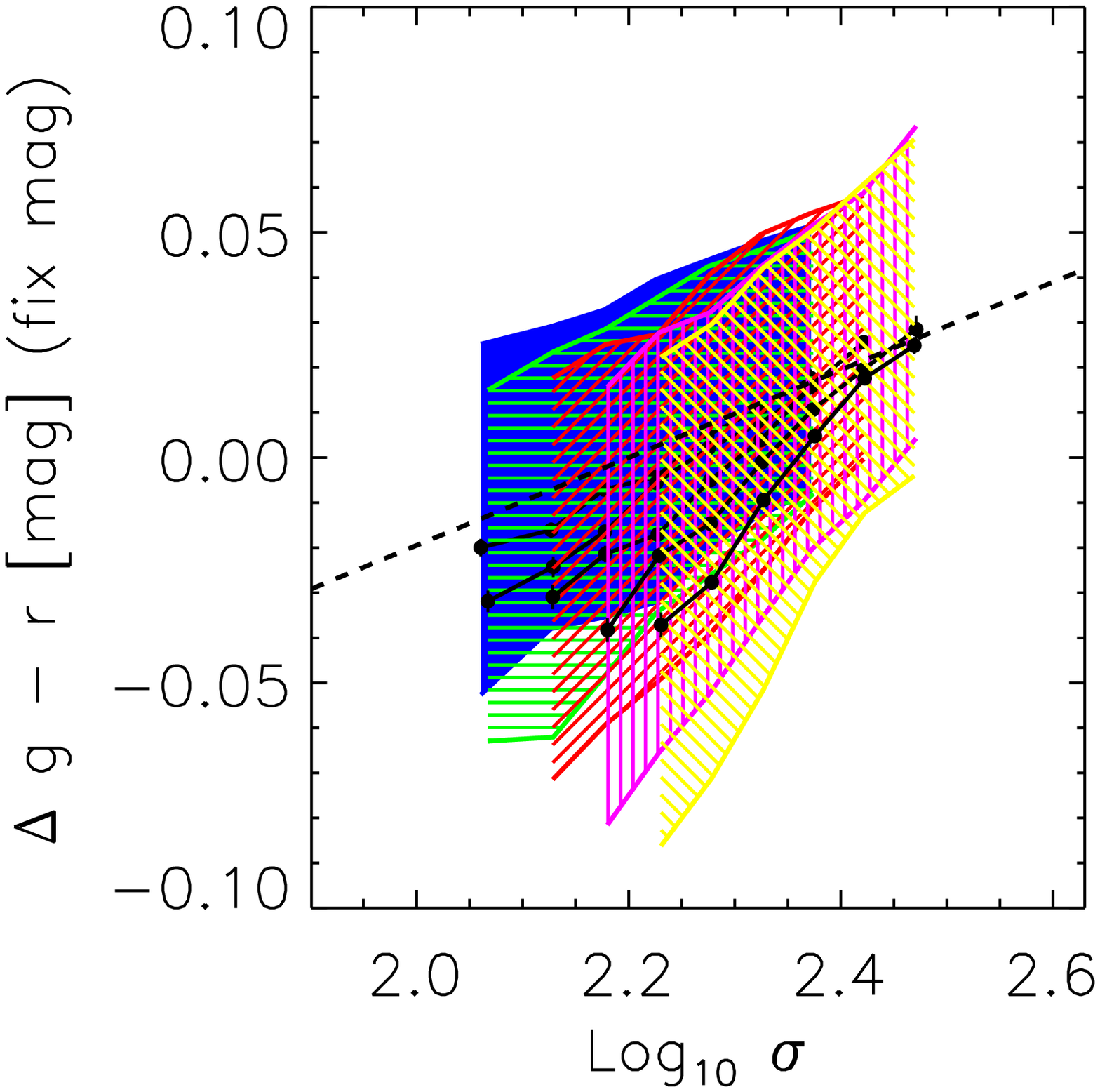}
 \caption{Residuals from the color-$\sigma$ relations do not correlate 
          with magnitude (upper panel), whereas residuals from the 
          color-magnitude relation do correlate with velocity dispersion 
          (bottom panel).  
          Dashed lines show the relations from our model,  
          equation~(\ref{basic}), obtained by inserting the values 
          from Tables~\ref{MLcov} and~\ref{MLcmag} in 
          equations~(\ref{RcvM}) and~(\ref{RcmV}), respectively.  
          The bottom panel can be interpreted as showing that galaxies 
          with large velocity dispersions tend to have older stellar 
          populations.  
          Although the steepening at higher redshifts in the bottom 
          panel suggests that the scatter around the mean color-magnitude 
          relation is larger at high redshift, or that evolution is  
          differential, it is, in fact, a selection effect (see text 
          between equations~\ref{RcmV} and~\ref{RcmRcv}).}
 \label{agesigma}
\end{figure}

Suppose that the color-magnitude relation is a consequence of the fact 
that metallicity increases with luminosity along the sequence, and that 
the galaxies which scatter redward of the relation are older.  
Then the correlation in Figure~\ref{agesigma} suggests that galaxies 
with large velocity dispersions are, on average, older.  

\begin{figure}[t]
 \centering
 \epsfxsize=0.98\hsize\epsffile{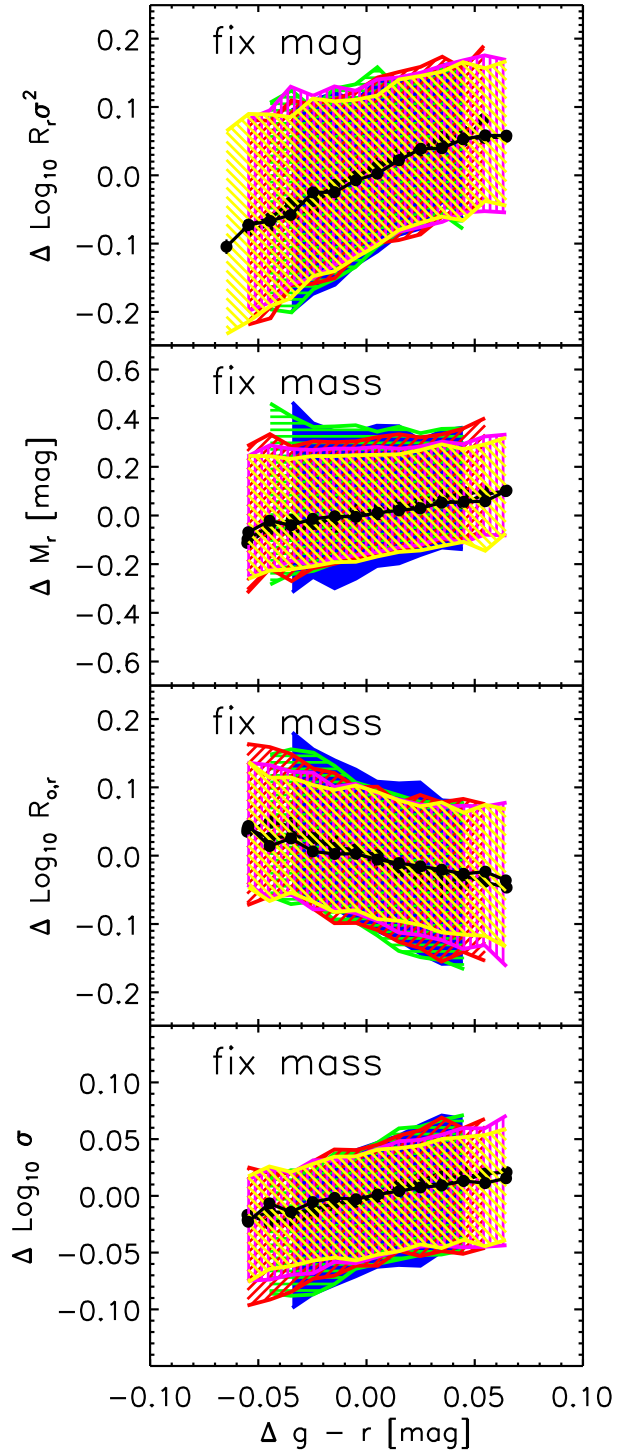}
 \caption{Galaxies which scatter blueward of the color-magnitude relation 
          scatter towards smaller masses from the mass-luminosty 
          relation (top).  At fixed mass, redder galaxies tend to be 
          fainter (second), have smaller sizes (third), and have larger 
          velocity dispersions (bottom).  
          Since mass $\propto R_r\sigma^2$, the trend with $\sigma$ 
          shown in the bottom panel could have been inferred directly 
          from the trend with size. }
 \label{massresid}
\end{figure}

The correlation in the panel on the left of Figure~\ref{cvmresids} 
implies that older galaxies tend to have larger velocity dispersions 
than their luminosities indicate.  A correlation between age and the 
residuals from the $\sigma$-luminosity relation was found by 
Forbes \& Ponman (1999), but their age estimates came from comparing  
the spectra of these objects with stellar population synthesis models.  
Our analysis indicates that age estimates from the color-magnitude 
relation lead to the same conclusion.  (Note that the analysis in 
Appendix~\ref{algebra} suggests that age is more closely related to 
the residuals in $\sigma$ with respect to the mean 
$\langle\sigma|L \rangle$ relation, than to the residuals in magnitude 
with respect to the $\langle L|\sigma\rangle$ relation.  That is, 
the residuals studied by Forbes \& Ponman (1999) should be more closely 
related to age than those studied more recently by Proctor et al. 2004.)

\begin{figure*}
 \centering
 \epsfxsize=\hsize\epsffile{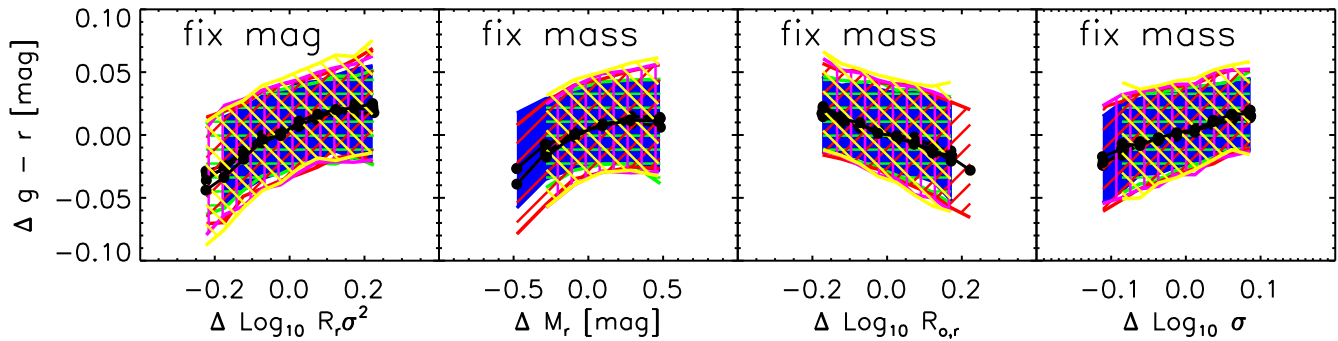}
 \caption{Similar to previous Figure, but now with color-magnitude 
          residual as the dependent parameter. }
 \label{massresid2}
\end{figure*}

Figure~\ref{lvrm} shows that luminosity is tightly correlated with 
mass $\propto R_r\sigma^2$.  
So it is interesting to study how residuals from the color-magnitude 
relation correlate with residuals from the mass-luminosity relation.  
The top-most panel of Figure~\ref{massresid} shows that galaxies 
which scatter blueward from the color-magnitude relation tend to 
scatter to smaller masses from the mass-luminosity relation.  
The other panels show that, at fixed mass, redder galaxies tend to 
be fainter (second from top), have smaller sizes (third from top), 
and have larger velocity dispersions (bottom).  Since luminosity and 
mass are so tightly correlated, we can think of the color-mass residual 
as being similar to the color-magnitude residual.  If we treat the 
color-magnitude residual as an age indicator, then Figure~\ref{massresid} 
indicates that, at fixed mass, older (redder) galaxies are fainter, 
smaller, and have larger velocity dispersions.  All of these trends are 
qualitatively similar to those of dark matter halos, but they are also 
consistent with a model in which the bluer light is less concentrated 
than the red light (see, e.g. Figure~3 in Bernardi et al. 2003a), 
as color gradients also suggest (e.g. Figure~7 in Bernardi et al. 2003d).  
Figure~\ref{massresid2} is similar, but now the color-magnitude residual 
is shown as the dependent parameter. 
 
In Figures~\ref{massresid} and~\ref{massresid2}, the first two panels 
from the top and left deserve further comment.  
The first panel indicates that bluer, lower mass galaxies can have the 
same luminosity as redder, more massive galaxies.  
This is easy to understand if the color deviation traces age.  
Suppose, instead, we wanted to construct a model in which the residuals 
from the color-magnitude relation trace metallicity rather than age, 
the redder galaxies being the more metal rich.  
In this case, one would interpret the first panel  as indicating 
that lower mass, metal poor galaxies can have the same luminosity as 
more massive, metal rich galaxies.  
In the Bruzual-Charlot (2003) models, at fixed mass, color changes 
are associated with larger magnitude changes if the color change 
is tied to age rather than metallicity (see top-left panel of 
Figure~\ref{bccmag}).  The two possiblities differ by a factor of 
two, so, in principle, one might to be able to distinguish between 
them using the second panels in Figures~\ref{massresid} 
and~\ref{massresid2}.  Figure~\ref{massresid} suggests that 
 $\langle\Delta M|\Delta(g-r)\rangle\propto \Delta(g-r)$, 
and Figure~\ref{massresid2} suggests 
 $\langle\Delta(g-r)|\Delta M\rangle\propto\Delta M/10$.  
These observed relations are shallower than both models:  
shallower relations indicate that scatter is important, 
but this makes it difficult to determine which of the two 
possibilities is more accurate.   

Appendix~\ref{algebra} develops in some detail the consequences of a 
model in which the assumptions that metallicity increases with 
luminosity along the color-magnitude relation, 
and that the residuals from the relation are indicators of age, 
are combined with our finding that the color-magnitude relation is 
determined by the color-$\sigma$ and magnitude-$\sigma$ relations.  
(The Appendix also shows that these conclusions are unchanged if 
we use the orthogonal rather than the direct fit, i.e., if we 
assume metallicity varies along the principal axis of the 
color-magnitude relation, and age varies perpendicular to this axis.)  
In such a model:  
\begin{itemize}
\item metallicity and $\sigma$ are correlated 
      (galaxies with large velocity dispersions tend to be more metal rich); 
\item age and $\sigma$ are correlated 
      (galaxies with large velocity dispersions tend to be older); 
\item lines which run parallel to the mean $\sigma$-magnitude relation 
      are approximately lines of constant age; 
\item in the plane of age along the $x$-axis and metallicity along the 
      $y$-axis, lines of constant velocity dispersion slope down and to 
      the right.  
\end{itemize}
If $\sigma$ is indeed sensitive to age, one might wonder if the slope 
of the color-$\sigma$ correlation steepens with redshift.  
The SDSS is a relatively shallow, magnitude-limited survey; the median 
redshift in our sample is of order 0.1, so the sample is not ideal for 
quantifying trends with redshift.  
On the other hand, it offers a large number of galaxies ($\sim 30000$) 
with well-calibrated surface photometry and velocity dispersions, 
so we may reasonably expect to be able to detect a small signal.  
We are also aided by the fact that, of the three pairwise relations, 
color-magnitude, $\sigma$-magnitude, and color-$\sigma$, it is the 
color-magnitude relation which is a consequence of the other two 
(c.f. Figures~\ref{cmag} and~\ref{csig}; also see Bernardi et al. 2003d).  
Because the color-$\sigma$ relation is the fundamental correlation, 
selection effects associated with the magnitude limit of our sample 
are unimportant (see Appendix~\ref{algebra} for details).  
The bottom left panel of Figure~\ref{csig} suggests that the slope 
of the color-$\sigma$ relation is steepening with redshift, but data 
from higher redshift and a better understanding of the appropriate 
$k+e$-corrections are required before we can claim to have measured 
this with confidence.  Other tantalizing evidence of evolution comes 
from Figure~\ref{agesigma}:  
the correlation between color-magnitude residuals (i.e., age) and 
velocity dispersion steepens systematically with redshift.  Although it 
is tempting to conclude that this is additional evidence of evolution, 
the analysis in Appendix~\ref{algebra} shows that it is, in fact, a 
selection effect.  

Taken together, the findings itemized above indicate that our data 
{\it require} the existence of an age-metallicity degeneracy; 
this degeneracy is most noticable for galaxies which have the same 
velocity dispersion $\sigma$.  In this respect, our findings are 
consistent with Figure~4 of Trager et al. (2000b).   
But our results are inconsistent with their conclusion that the most 
massive galaxies are less metal rich (see their Figure~6).  
Indeed, Figure~\ref{bccmag} indicates that it would be difficult for 
single burst models to produce the observed color-magnitude relation 
if the more luminous galaxies were older but less metal rich.  
Our finding that older galaxies tend to have larger velocity 
dispersions than their luminosities indicate is qualitatively 
consistent with one of the conclusions of Forbes \& Ponman (1999); 
this is encouraging because their age estimates come from a very 
different approach (detailed comparison of spectral features with 
stellar population synthesis models).  

While it is tempting to calibrate the relation between 
color-magnitude residual and age, this would be model-dependent.  
For example, the Bruzual-Charlot models presented in 
Appendix~\ref{bc2003} can be used to perform this calibration.  
But we are hesitant to take this path because the models do not include an 
important correlation between velocity dispersion and $\alpha$-element 
abundance ratios (Trager et al. 2000a; Thomas, Maraston \& Bender 2003; 
Proctor et al. 2004).  

In the model developed here, the scatter around the color-magnitude 
and the $\sigma$-magnitude relations should increase with lookback 
time.  This can be tested with data from redshifts of order one-half 
to unity which will soon be available (e.g. the SDSS-2dF survey of 
luminous red galaxies at
http://sdss2df.ncsa.uiuc.edu/).  

\bigskip

MB thanks A. Connolly for support.  

Funding for the creation and distribution of the SDSS Archive has
been provided by the Alfred P. Sloan Foundation, the Participating
Institutions, the National Aeronautics and Space Administration, the
National Science Foundation, the U.S. Department of Energy, the
Japanese Monbukagakusho, and the Max Planck Society. The SDSS Web site
is http://www.sdss.org/. 

The SDSS is managed by the Astrophysical Research Consortium (ARC)
for the Participating Institutions. The Participating Institutions are
The University of Chicago, Fermilab, the Institute for Advanced Study,
the Japan Participation Group, The Johns Hopkins University, 
the Korean Scientist Group, Los Alamos National Laboratory, 
the Max-Planck-Institute for Astronomy (MPIA),  
the Max-Planck-Institute for Astrophysics (MPA), 
New Mexico State University, University of Pittsburgh, 
Princeton University, the United States Naval Observatory,
and the University of Washington.

{}

\appendix

\section{A:  Single-burst models}\label{bc2003}

In the single-burst models of Bruzual \& Charlot (2003), galaxy colors 
depend on both age and metallicity.  For a population which is older 
than one Gyr, and has solar metallicity or greater, we have found that 
the color is quite well described by:  
\begin{eqnarray*}
 g-r &=& 0.7 + 0.25(T-9.5) + {0.15\over 0.40}\,Z \qquad {\rm if\ T>9.5}
  \nonumber\\
 g-r &=& 0.7 + 0.50(T-9.5) + {0.15\over 0.40}\,Z \qquad {\rm if\ T<9.5}
\end{eqnarray*}
where $T = \log_{10}$(age/Gyr).  

The four panels of Figure~\ref{bccmag} provide different examples of 
the color-magnitude relation associated with these models.  
The top left panel shows the locus traced out in the color-magnitude 
plane by a single object of fixed mass as it ages.  The two dotted 
lines which slope from top left to bottom right show lines of constant 
metallicity (higher metallicity is redder), and the approximately 
vertical lines which connect them show locii of constant age (older is 
fainter).  This panel is relevant to results presented in 
Figures~\ref{massresid} and~\ref{massresid2}.  It shows that, at fixed 
mass, color changes are associated with larger magnitude changes if the 
color change is tied to age rather than metallicity:  
$\Delta M\propto 5\,\Delta(g-r)$ if the color change is entirely due to 
age, whereas $\Delta M\propto (5/2)\,\Delta(g-r)$ if the color change 
is entirely due to metallicity.  

\setcounter{figure}{0}

\begin{figure*}[t]
 \centering
 \epsfxsize=0.75\hsize\epsffile{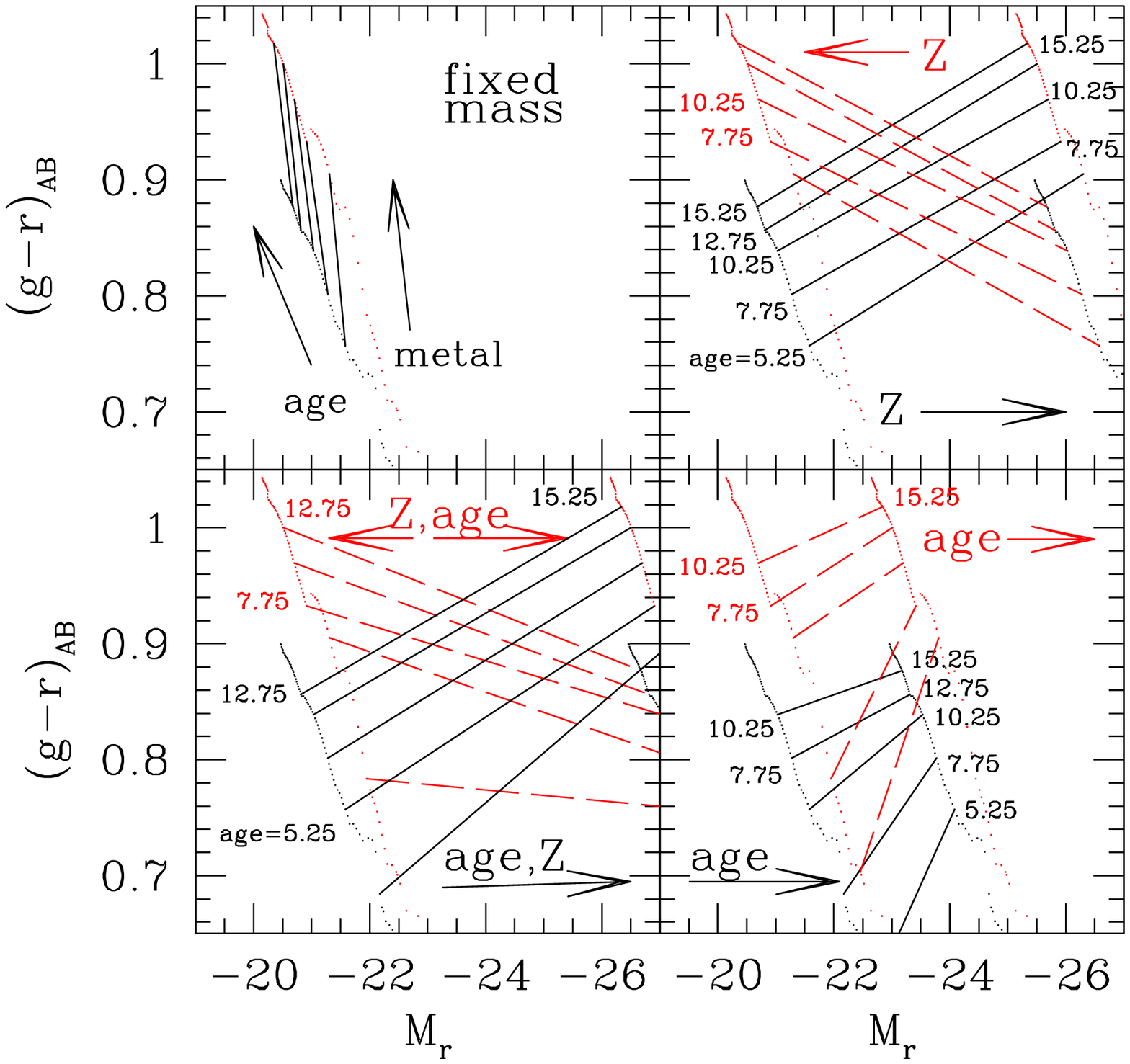}
 \caption{Correlation between color and magnitude in the single-burst 
          models of Bruzual \& Charlot (2003).  Different panels show 
          the result of assuming the various correlations between 
          luminosity, age and metallicity that are discussed in the 
          main text.  
          The top right panel shows color magnitude relations in 
          models which have metallicity varying with luminosity.  
          The bottom right panel shows relations in which age 
          increases with luminosity.  Bottom left panel shows 
          models in which age and metallicity vary with luminosity.  
          In the top two panels, solid and dashed lines which 
          connect dots trace locii of constant age = 15.25, 12.75, 
          10.25, 7.75 and 5.25~Gyrs, with the oldest being reddest.}
\label{bccmag}
\end{figure*}

Top right panel shows models in which metallicity correlates with 
luminosity.  The solid lines which span approximately the entire 
luminosity range show locii of constant age in a model which has 
metallicity increasing with luminosity.  The dashed lines show 
constant age in a model where metallicity decreases with luminosity.  
Notice that, in both cases, the slope of the color magnitude relation 
is approximately independent of age.  However, the data indicate 
that color increases with luminosity, so that models with metallicity 
decreasing with luminosity are not permitted.  

The bottom right panel shows two models in which age, rather than 
metallicity, vary along the luminosity axis.  Both models assume that 
older objects are more luminous, but the redder set of lines (i.e., 
the dashed lines) show a more metal rich population.  In both cases, 
the slope of the relation varies more strongly with age than it did in 
the previous panel, with the slope of the color-magnitude relation 
evolving rapidly when the population is younger, but less rapidly at later 
times.  

The two panels on the right showed models in which either metallicity 
or age varied with luminosity, but not both.  
Dashed lines which slope down and to the right in the bottom left panel 
show the color-magnitude relation  in a model where the more luminous 
objects are older but metal poor.  This model is qualitatively like 
those of Trager et al. (2000b).  
Solid lines show a model in which both age and metallicity increase 
with luminosity.  The data are more consistent with this latter model 
which has more luminous objects being older and metal rich.

\section{B: Luminosities, colors, velocity dispersions, ages and metallicities}
\label{algebra}
\subsection{Luminosity, color and velocity dispersion}
Define $m = (M-M_*)/\sigma_{MM}$, $v = (V-V_*)/\sigma_{VV}$, 
and $c = (C-C_{*})/\sigma_{CC}$, 
where $M$ is the absolute magnitude, 
$V\equiv\log_{10}$(velocity dispersion/km~s$^{-1}$), 
and $C$ denotes a color.  Our notation uses $X_*$ amd $\sigma_{XX}$ to 
denote the mean and rms of observable $X=M,V,C$, etc.  
This Appendix considers a model in which the only 
fundamental correlations are with $\sigma$:    
\begin{equation}
 p(v,m,c) = p(v)\,p(m|v)\,p(c|v).  
 \label{basic}
\end{equation}
Pairwise correlations, such as that between luminosity and $\sigma$, 
are given by integrals like 
\begin{equation}
 \langle mv\rangle = \int dm\, dv\, dc\,(mv)\,p(v,m,c).
\end{equation}
For the assumed model, these are relatively simple to evaluate.  
If we suppose that the mean magnitude at fixed $\sigma$ scales as 
$\langle m|v\rangle = \xi_{mv}v$, and the mean color at fixed $\sigma$ 
is also linear in $v$, say $\xi_{cv}v$, then 
\begin{equation}
 \langle mv\rangle = \xi_{mv}, \ {\rm and}\ \langle cv\rangle = \xi_{cv}.  
\end{equation}
This comes from our choice to normalize $\langle vv\rangle = 1$, 
and similarly for the other variables.  

Let $\sigma_{m|v}$ denote the scatter in $m$ around $\langle m|v\rangle$.  
If $\sigma_{m|v}$ is independent of $v$, then 
\begin{eqnarray}
 \langle v^2\rangle &=& \int dm\, p(m)\, \langle v^2|m\rangle \nonumber\\
                &=& \int dm\, p(m)\, [\sigma_{v|m}^2 + \xi_{vm}^2m^2]\nonumber\\
                &=& \sigma_{v|m}^2 + \xi_{vm}^2
\end{eqnarray}
Therefore, $\sigma_{m|v}^2 = 1 - \xi_{mv}^2$, and similarly for the scatter 
around the color-$\sigma$ relation.  

What if we are interested in relations which do not involve $\sigma$?  
For instance, the mean color at fixed magnitude is 
\begin{eqnarray}
 \langle c|m\rangle &=& \int dv dc\,c\,p(c|v)\, p(v|m) \nonumber\\
                      &=& \int dv \xi_{cv}v\, p(v|m) = \xi_{cv}\xi_{mv}\,m.
\end{eqnarray}
This shows that the slope of the color-magnitude relation can be 
written in terms of the slopes of the other two relations:  
$\xi_{cm} = \xi_{cv}\xi_{mv}$.  This is a simple approximation to 
the observed correlations: $\xi_{cm}=-0.36$ and $\xi_{cv}\xi_{mv}=-0.40$ 
(from Tables~\ref{MLcov} and~\ref{MLcmag}).   

By assumption~(\ref{basic}), the residuals of the $m$-$\sigma$ relation, 
$R_{m|v}=m-\langle m|v\rangle = m-\xi_{mv}v$, 
do not correlate with those of the color-$\sigma$ relation.  
It is straightforward to verify that this is so:  
\begin{eqnarray}
 \langle R_{m|v}R_{c|v}\rangle 
 &=& \langle(m-\xi_{mv}v)(c-\xi_{cv}v)\rangle \nonumber\\ 
 &=& \langle mc - \xi_{cv}mv -\xi_{mv}vc + \xi_{mv}\xi_{cv}v^2\rangle 
      \nonumber\\
 %&=& \xi_{cm} - \xi_{cv}\xi_{mv} - \xi_{mv}\xi_{cv} + \xi_{mv}\xi_{cv} 
 % \nonumber\\
 &=& 0.
 \label{RmvRcv}
\end{eqnarray}
However, the residuals of the color-magnitude relations do correlate 
with those of the $\sigma$-magnitude relation:
\begin{eqnarray}
 \langle R_{c|m}R_{v|m}\rangle 
 &\equiv& \langle(c-\xi_{cm}m)(v-\xi_{vm}m)\rangle \nonumber\\ 
 &=& \langle cv - \xi_{cm}mv - \xi_{vm}mc + \xi_{cm}\xi_{vm}m^2\rangle 
     \nonumber\\
 &=& \xi_{cv} - \xi_{cm}\xi_{mv} = \xi_{cv} \, (1 - \xi_{mv}^2).
 \label{RcmRvm}
\end{eqnarray}
Thus, this model explains the correlations between residuals shown in 
Figure~\ref{cvmresids}.  

Assumption~(\ref{basic}) has another important consequence.  Namely, 
the slope of the color-$\sigma$ relation, at fixed magnitude, is 
\begin{equation}
 \langle c|v,m\rangle = {\int dc\,c\,p(c|v)\,p(m|v)\,p(v)\over p(m,v)} 
                      = \xi_{cv}v.  
\end{equation}
This shows that the slope of the $c$-$v$ correlation is the same as 
when the constraint on magnitudes is removed.  
Therefore, the slope of the color-$\sigma$ correlation measured in 
a magnitude limited sample is the same as the slope of the intrinsic 
relation.  In other words, if the only correlation between color and 
magnitude is through $\sigma$, then the slope of the color-$\sigma$ 
relation measured in a magnitude limited sample is not biased by 
selection effects.  We use this in the main text.  

\subsection{Age, metallicity and velocity dispersion}
If the residuals from the color-magnitude relation, $R_{c|m}$, are 
indicators of age, and if the relation between age and $R_{c|m}$ is 
linear, then the mean age as a function of velocity dispersion can be 
obtained from 
\begin{equation}
 \langle R_{c|m}|v\rangle = 
 %   (<c|v>-<c|m>) p(m|v) = (\xi_{cv} - \xi_{cm}\xi_{mv})\,v = 
                            (1 - \xi_{mv}^2)\,\xi_{cv}\,v.
 \label{RcmV}
\end{equation}
If $\xi_{cv}>0$, then age increases with increasing velocity dispersion.  
The rms scatter around this relation is $\sigma_{c|v}$.  
The bottom panel of Figure~\ref{agesigma} shows this relation in the data.  
In this model, the mean age increases less rapidly with velocity dispersion 
(by a factor $1 - \xi_{mv}^2$) than does the color, suggesting that some 
of the increase in color is associated with factors other than age.  

In Figure~\ref{agesigma}, the relation between age and $\sigma$ appears 
to steepen with redshift.  If real, what does this steepening imply?  
Note that the quantities in the Figure have not been normalized 
by their rms.  If we suppose that the correlations between velocity 
dispersion, color, and magnitude do not evolve, but their rms values 
do, then the steepening implies that 
$(\sigma_{CC}/\sigma_{VV})$ is increasing.  The distribution of velocity 
dispersions does not appear to be evolving (e.g. Sheth et al. 2003), 
and there are no strong reasons to expect it to.  On the other hand, 
we argued in the Introduction that one expects to see the scatter around 
the color-magnitude relation increase with redshift.  Although we do 
expect to see $\sigma_{CC}$ increase, the increase implied by 
Figure~\ref{agesigma} is larger than expected.  This is because the 
steepening is a selection effect---it is the result of the magnitude 
limit of the survey.  To see why, note that the magnitude limit restricts 
the range in $m$ over which the integrals which define 
$\langle R_{c|m}|v\rangle = \int dm\,p(m|v)\,(\xi_{cv} v - \xi_{cm} m)$
are performed.  If we write the result as 
$(1 - \xi_{mv}^2)\,\xi_{cv}v$ times a correction factor 
$\int dm\,p(m|v)\,(\xi_{cv} v - \xi_{cm} m)/(\xi_{cv} v - \xi_{cm}\xi_{mv}v)$, 
then this factor depends on the magnitude limit, so in effect, 
the slope of the correlation with $v$ depends on redshift.  

\setcounter{figure}{0}

\begin{figure*}
 \centering
 \epsfxsize=0.45\hsize\epsffile{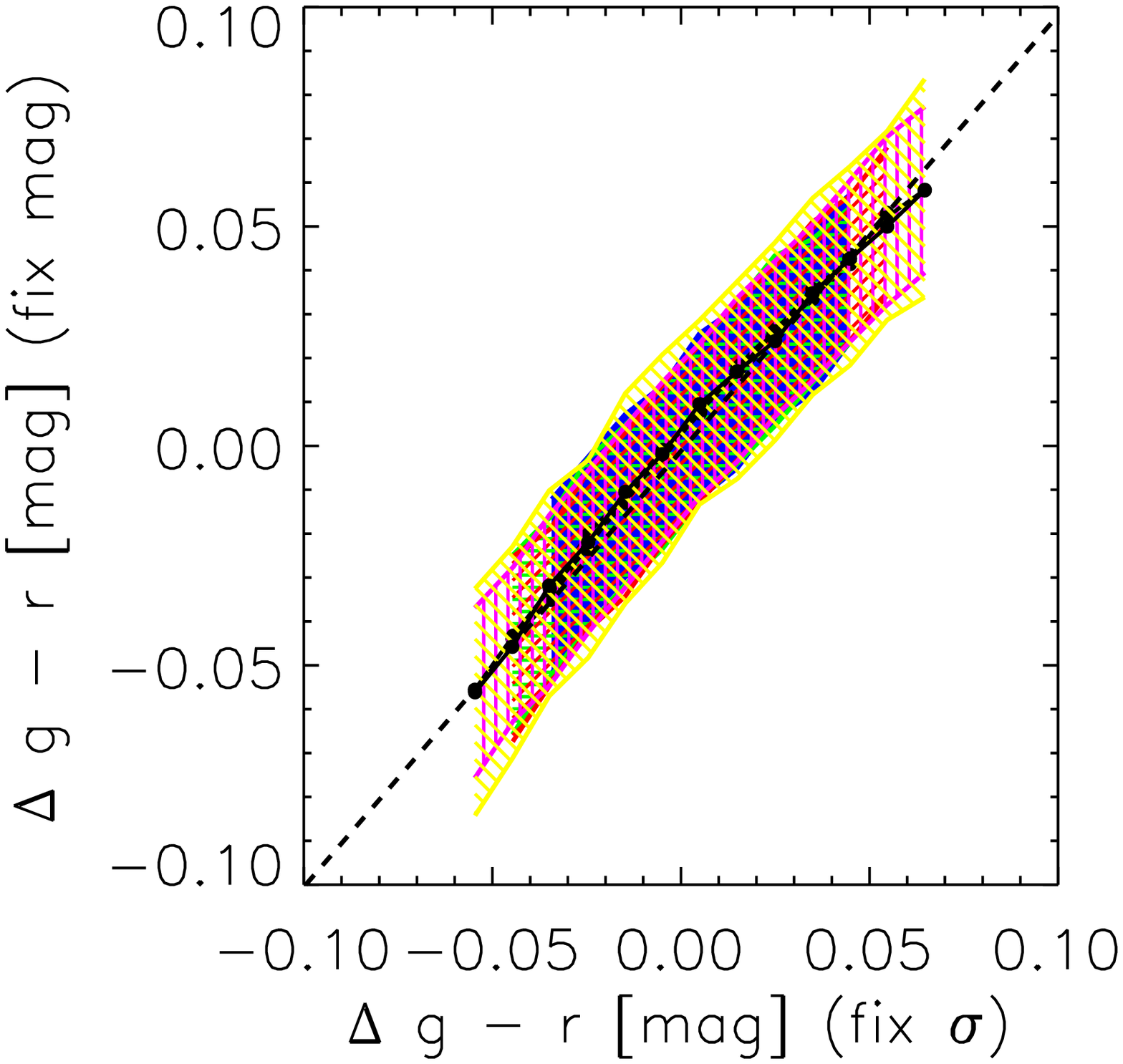}
 \epsfxsize=0.45\hsize\epsffile{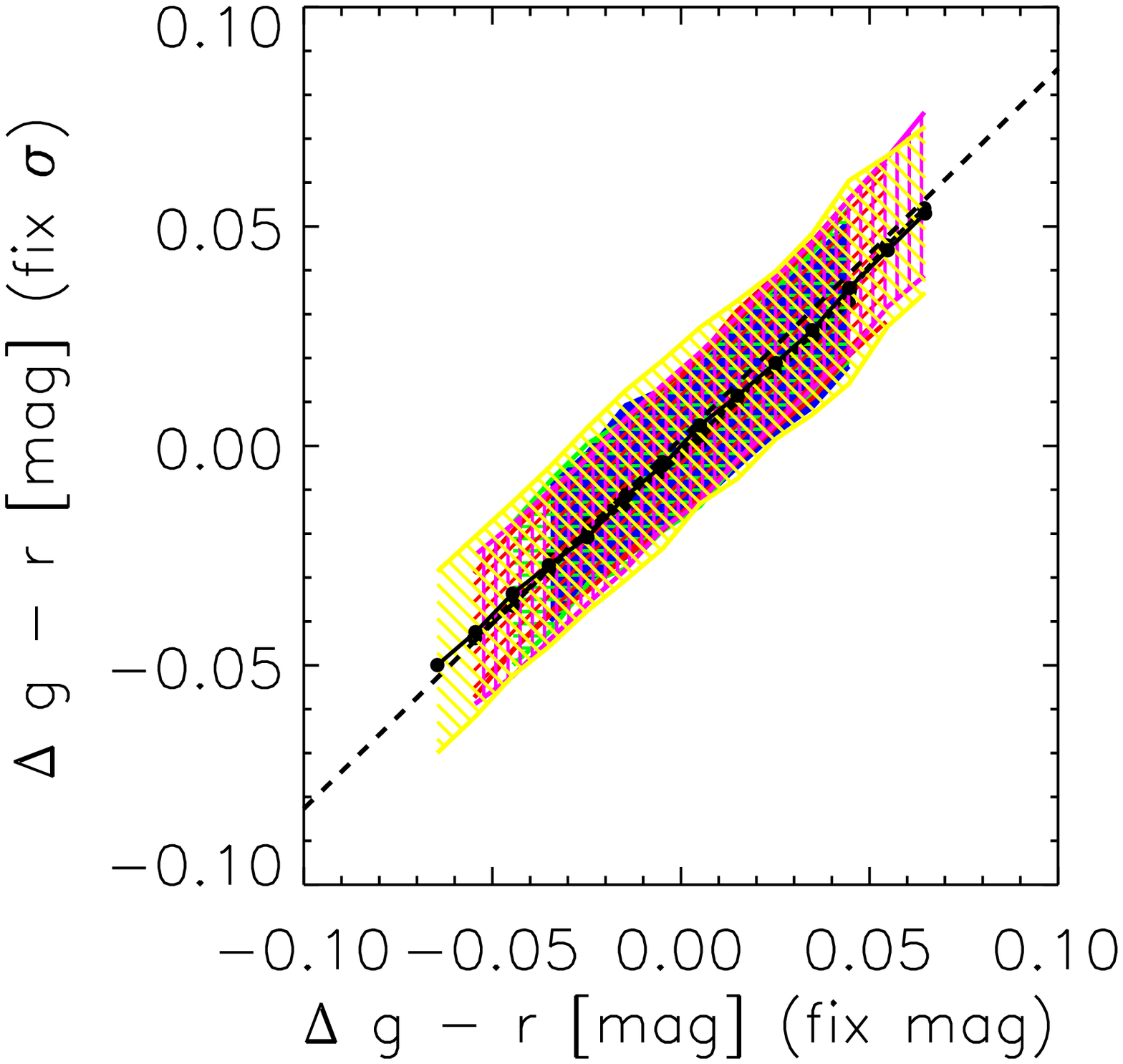}
 \caption{Correlations between residuals from the color-magnitude 
          and color-$\sigma$ relations are well described by the model 
          developed in the text.  Dashed line in panel on the left shows 
          the correlation described by equation~(\ref{RcmRcv}), and the 
          line in the panel on the right shows equation~(\ref{RcvRcm}).  }
 \label{rCVrCM}
\end{figure*}

If the residuals from the color-$\sigma$ relation are a consequence 
of age and possibly other variables (such as metallicity and/or 
environment), then correlations between $R_{c|m}$ and $R_{c|v}$ 
can be used to study how much of the scatter in the color-$\sigma$ 
relation is due to age, and how much to other effects.  For instance, 
\begin{eqnarray}
 \label{RcmRcv}
 \langle R_{c|m}|R_{c|v}\rangle &=& R_{c|v}, \quad {\rm and}\\
 \langle R_{c|v}|R_{c|m}\rangle &=& 
    R_{c|m}\,{\sigma_{c|v}^2\over\sigma_{c|m}^2} = 
    R_{c|m}\,{1 - \xi_{cv}^2\over 1 - \xi_{cv}^2\xi_{mv}^2}.
 \label{RcvRcm}
\end{eqnarray}
This is consistent with the data has shown in Figure~\ref{rCVrCM}.
Thus, although objects which have the same fixed offset in color from 
the color-$\sigma$ relation may have a range of ages, the mean age is 
the same as that infered from the color-offset of the color-magnitude 
diagram (see upper panel of Figure~\ref{rCVrCM}).  
On the other hand, the mean color offset from the 
color-$\sigma$ relation of objects which have the same age (i.e., they 
are all offset from the color-magnitude relation by the same amount) is 
smaller than the color-offset from the color-magnitude relation 
(see bottom panel of Figure~\ref{rCVrCM}).  
This is because some the color change 
associated with age effects has been accounted-for by the fact that age 
varies with $\sigma$.  

In contrast, the result from Figure~\ref{agesigma} (upper panel) showing 
\begin{equation}
 \langle R_{c|v}|m\rangle = 0
 \label{RcvM}
\end{equation}
is less straightforward to interpret.  
This is because the previous correlations suggest that 
residuals from the color-$\sigma$ relation are some combination of 
both age and metallicity.  If luminosity is driven primarily by 
metallicity, then the fact that $\langle R_{c|v}|m\rangle = 0$ must 
mean that the age effects exactly cancel the correlation with 
metallicity.  Moreover, if we replace luminosity with metallicity, 
then the fact that $\langle R_{c|m}|m\rangle = 0$ suggests there is 
no correlation between age and metallicity.  

The mean age at fixed luminosity and velocity dispersion is 
\begin{equation}
 \langle R_{c|m}|v,m\rangle = \xi_{cv}(v - \xi_{mv}m),
 \label{agevm}
\end{equation}
and the rms scatter around this mean is $\sigma_{c|v}$.  
Since the final expression is really  $\xi_{cv}\,R_{v|m}$, 
this shows that objects which have larger velocity dispersions than 
the mean $\sigma$-magnitude relation tend to scatter redward of the 
color-magnitude relation, and so are, on average, older than those 
which scatter to smaller dispersions.  This has two important consequences.  
First, lines which run parallel to the mean $\langle v|m\rangle$ 
relation trace out curves of approximately constant age, and this 
statement becomes increasingly accurate if the relation between color 
and $\sigma$ is tight (recall that the scatter around 
$\langle R_{c|m}|v,m\rangle$ is $\sigma_{c|v}$).  
Second, if younger objects evolve more rapidly, and the evolution is 
stronger in luminosity than in velocity dispersion, then the scatter 
around the $\sigma$-magnitude relation should be larger at high redshift.  

The mean velocity dispersion at fixed magnitude and color-magnitude 
residual is 
\begin{equation}
 \langle v|m,R_{c|m}\rangle = \xi_{mv}\,m + 
                              {\xi_{cv}(1-\xi_{mv}^2)\over 1-\xi_{cm}^2}\,R_{c|m},
\end{equation}
with rms scatter 
% the square-root of $(1-\xi_{cv}^2)(1-\xi_{mv}^2)/(1 - \xi_{mv}^2\xi_{cv}^2)$.  
$\sigma_{v|m}\sigma_{c|v}/\sigma_{c|m}$.  

If age and metallicity scale as $T=aR_{c|m}$ and $Z=-bm$, with 
$a$ and $b$ both positive, then the same average value of $\sigma$ 
can be obtained by increasing the age and decreasing the metallicity, 
or vice-versa (we have assumed that $\xi_{mv}<0$, since, 
as Figure~\ref{lvrm} shows, velocities tend to increase with luminosity).  
We have not discussed the roles played by the environment and by 
changes in $\alpha$-element abundances.  
In this model, they would be responsible for some of the scatter 
around these mean trends with age and metallicity.  

The expression above suggests that different combinations of age and 
metallicity can yield the same average age.  We can see this more 
directly by computing the correlation between age and metallicity at 
fixed velocity dispersion:
\begin{equation}
 \langle m R_{c|m}|v\rangle - \langle m|v\rangle \langle R_{c|m}|v\rangle = 
 -\xi_{cm}\sigma^2_{m|v}
\end{equation}
If age and metallicity scale as $T=aR_{c|m}$ and $Z=-bm$, with 
$a$ and $b$ both positive and $\xi_{cm}<0$, then the same value of 
$\sigma$ can be obtained by increasing the age and decreasing the 
metallicity, or vice-versa.  This is also consistent with 
equation~(\ref{agevm}):  
the mean age at fixed velocity dispersion increases as the metallicity 
decreases.  

So far, we have not asked how one calibrates the relation between 
color-magnitude residual and age.  In single burst models, it is 
possible to constrain the ratio $a/b$, because, at fixed $\sigma$, 
log$_{10}$(age) and metallicity are expected to be related by a 
factor of $-3/2$ (e.g. Worthey 1994).  

\medskip

\noindent{\bf In summary:}
We showed that the color-magnitude relation is determined by the 
color-$\sigma$ and magnitude-$\sigma$ relations.  
Previous work suggests that 
 (i) metallicity increases with luminosity along the color-magnitude 
   relation (as suggested by the observation that the slope of the 
   color-magnitude relation does not evolve), and 
 (ii) that the residuals from the relation are indicators of age.  
We developed a simple model which combines these results.  
The model implies that:  
 (i) age and $\sigma$ are correlated (galaxies with large velocity 
      dispersions tend to be older); 
 (ii) lines which run parallel to the mean $\sigma$-magnitude relation 
      are approximately lines of constant age; 
 (iii) in the plane of age along the x-axis and metallicity along the 
      y-axis, lines of constant velocity dispersion slope down and 
      to the right;
 (iv) the scatter around the color-magnitude and the $\sigma$-magnitude 
      relations should increase with lookback time.  

\subsection{The orthogonal fit}
The previous section developed a model for correlations between 
age and $\sigma$ in which luminosity was proportional to metallicity, 
and residuals from the color-magnitude relation were proportional to 
log(age).  
Here, we change these assumptions slightly, and ask what happens if 
metallicity varies along the long axis of the color-magnitude relation, 
whereas age varies along the short axis.  
In this case, the age-metallicity plane is simply a rotation of the 
color-magnitude plane, so metallicity is some combination of luminosity 
and color, and age is some different combination.  In particular, 
this would set 
\begin{equation}
 T \propto {\sigma_c c - a_0 \sigma_m m\over\sqrt{1 + a_0^2}}
   \quad{\rm and}\quad 
 Z \propto -{a_0 \sigma_c c + \sigma_m m\over\sqrt{1 + a_0^2}},
\end{equation}
where 
\begin{displaymath}
 a_0 = {\sigma^2_c - \sigma_m^2\over 2\xi_{cm}\sigma_c\sigma_m} + 
 \sqrt{1+\left({\sigma^2_c -\sigma_m^2\over 2\xi_{cm}\sigma_c\sigma_m}\right)^2}.
\end{displaymath}
When $\sigma_c\ll \sigma_m$ then $a_0\to \xi_{cm}\sigma_c/\sigma_m$, 
the mean age $T$ at fixed velocity dispersion is
 $\propto \sigma_c (1 - \xi_{mv}^2)\,\xi_{cv}\,v$, 
and the mean value of $Z$ at fixed $v$ is 
$\propto -v \sigma_m \xi_{mv} (1 + \xi_{cv}^2\sigma_c^2/\sigma_m^2)/\sqrt{1+a_0^2} 
 \to -v \sigma_m \xi_{mv}$.  
Thus, in this limit (because the luminosity in different bands is 
quite well correlated, the data do indeed have $\sigma_c\ll\sigma_m$), 
analysis of the orthogonal fit produces results which are similar to 
the analysis of the previous section.  

If the scatter in colors were larger, the difference between the 
orthogonal fit model and the one studied in the previous subsection 
would also be larger.  
% $\langle Z|m\rangle \propto -m \sigma_m \sqrt{1 + a_0^2}$,
For instance, if $\sigma_m=\sigma_c$, then the age indicator would 
be $\propto (c - m)/\sqrt{2}$ and the metallicity would be 
$\propto (-c - m)/\sqrt{2}$, so the mean age at fixed velocity dispersion 
would be $\propto v\,(\xi_{cv} - \xi_{mv})/\sqrt{2}$, 
whereas the mean metallicity at fixed $v$ would scale as 
$-v\,(\xi_{cv} + \xi_{mv})/\sqrt{2}$.  
The rms scatter around the age-$\sigma$ and metallicity-$\sigma$ 
relations would be $\sqrt{(\sigma_{m|v}^2 + \sigma_{c|v}^2)/2}$.  
%<(c - m - <c|v> + <m|v>)^2|v>/2 = (\sigma_{c|v}^2 + \sigma_{m|v}^2)/2
%$\sqrt{1 - (\xi_{mv}^2+\xi_{cv}^2)/2}$.
The mean correlation between age and metallicity at fixed $v$ is 
% <(c-m)(c+m)|v> = <c^2 - m^2|v> = sig_cv^2 - sig_mv^2 = 1-ccv^2 - (1-cmv^2)
$(\xi_{cv}^2-\xi_{mv}^2)/2$.  
In the sample, $\xi_{mv}<0$ and $|\xi_{mv}|>\xi_{cv}$, 
so the mean age and metallicity increase with velocity dispersion, 
and if the age is larger than expected for the velocity dispersion, 
then the metallicity is smaller.  
Moreover, the mean age would increase as luminosity increases:
% <c-m|m> = (ccm - cmm) m
$(\xi_{cm}-1)\,m/\sqrt{2} = -m\,(1-\xi_{cv}\xi_{mv})/\sqrt{2}$, 
with rms scatter $\sigma_{c|m}$,  
and the mean metallicity would increase as 
% <c+m|m> = (ccm + cmm) m
$\langle Z|m\rangle = -m\,(1 + \xi_{cv}\xi_{mv})/\sqrt{2}$, with the same 
rms scatter, $\sigma_{c|m}$.  
Although some of these conclusions are qualitatively similar to those 
of the previous subsection, they are generally quantitatively different:  
the correlations between age and velocity dispersion and age and luminosity 
become steeper, whereas those between metallicity and $\sigma$ and 
metallicity and luminosity become shallower.

\end{document}